\documentclass[10pt,twocolumn,letterpaper]{article}

\usepackage{cvpr}
\usepackage{times}
\usepackage{epsfig}
\usepackage{graphicx}
\usepackage{amsmath}
\usepackage{amssymb}
\usepackage{tabularx}
\usepackage{xcolor}

\usepackage{kotex}
\usepackage{amsmath,amssymb}
\DeclareMathOperator{\E}{\mathbb{E}}
\usepackage{subcaption}

\usepackage[breaklinks=true,bookmarks=false]{hyperref}

\cvprfinalcopy % *** Uncomment this line for the final submission

\def\cvprPaperID{****} % *** Enter the CVPR Paper ID here

\ifcvprfinal\fi
\begin{document}

%%%%%%%%% TITLE
\title{GRDN:Grouped Residual Dense Network for Real Image Denoising and GAN-based Real-world Noise Modeling}

\author{Dong-Wook Kim\thanks{These authors contributed equally. Corresponding author: S.-W. Jung} , Jae Ryun Chung$^*$, and Seung-Won Jung\\
Department of Multimedia Engineering\\
Dongguk University, 04620, Seoul, Korea\\
{\tt\small spnova12@gmail.com, wjdwofus1004@gmail.com, swjung83@gmail.com}
}

\maketitle
\thispagestyle{empty}

\begin{figure*}[!t]
	\begin{center}
		\includegraphics[width=0.9\linewidth]{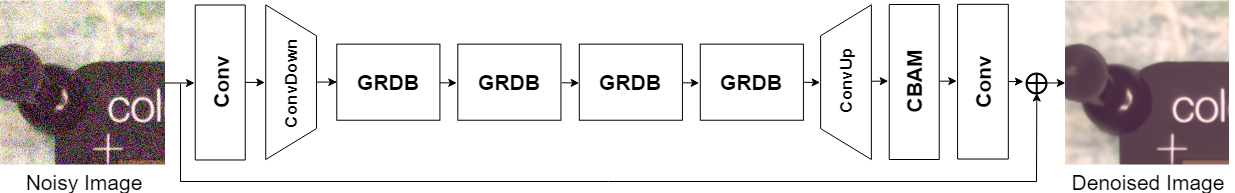}
	\end{center}
	\caption{The proposed network architecture: GRDN.}
	\label{fig:GRDN}
\end{figure*}

%%%%%%%%% ABSTRACT
\begin{abstract}
Recent research on image denoising has progressed with the development of deep learning architectures, especially convolutional neural networks. However, real-world image denoising is still very challenging because it is not possible to obtain ideal pairs of ground-truth images and real-world noisy images. Owing to the recent release of benchmark datasets, the interest of the image denoising community is now moving toward the real-world denoising problem. In this paper, we propose a grouped residual dense network (GRDN), which is an extended and generalized architecture of the state-of-the-art residual dense network (RDN). The core part of RDN is defined as grouped residual dense block (GRDB) and used as a building module of GRDN. We experimentally show that the image denoising performance can be significantly improved by cascading GRDBs. In addition to the network architecture design, we also develop a new generative adversarial network-based real-world noise modeling method. We demonstrate the superiority of the proposed methods by achieving the highest score in terms of both the peak signal-to-noise ratio and the structural similarity in the NTIRE2019 Real Image Denoising Challenge - Track 2:sRGB.
\end{abstract}

%%%%%%%%% BODY TEXT\part{title}

\section{Introduction}
In the field of image denoising, recent studies show that learning-based methods are more efficient than previous handcrafted methods such as block matching 3D (BM3D)~\cite{bm3d} and its variants. It is essential for learning-based methods to have a sufficient amount of dataset with high quality. Because a pair of noisy and noise-free images can be easily constructed by adding synthetic noise to noise-free images, a majority of previous learning-based methods focus on the classic Gaussian denoising task and pay the most attention to the architecture design of networks, especially convolutional neural networks (CNNs). However, due to the gap between synthetically generated noisy images and real-world noisy images, it was found that CNNs trained using synthetic images do not perform well on real-world noisy images and sometimes even inferior to BM3D~\cite{dnd}.

Toward real-world image denoising, there have been two main approaches. The first approach is to find a better statistical model of real-world noise rather than the additive white Gaussian noise~\cite{google_denoier,poissonian-gaussian,our_synthetic_dataset,poissonian2,n3net}. In particular, a combination of Gaussian and Poisson distributions was shown to closely model both signal-dependent and signal-independent noise. The networks trained using these new synthetic noisy images demonstrated the superiority in denoising real-world noisy images. One clear advantage of this approach is that we can have infinitely many training image pairs by simply adding the synthetic noise to noise-free ground-truth images. However, it is still arguable whether the real-world noise can be modeled by statistic models. The second approach is thus in an opposite direction. From real-world noisy images, nearly noise-free ground-truth images can be obtained by inverting an image acquisition procedure~\cite{SIDD,seeindark,towardlearning_pipline, dnd, RENOIR}. To our knowledge, smartphone image denoising dataset (SIDD)~\cite{SIDD} is one of the largest high quality image datasets on the second approach. However, the amount of provided images may not be enough for training a large network and without a sufficient knowhow it is difficult to generate ground-truth images from real-world noisy images. We thus adopt the second approach but applied our own generative adversarial network (GAN)-based data augmentation technique to obtain a larger dataset.

The network architecture is of course the utmost important. In CNN-based image restoration, dense residual blocks (RDBs)~\cite{rdn,rdn2} have received great attention. In this paper, we propose a new architecture called grouped residual dense network (GRDN). In particular, the proposed architecture adopts the recent residual dense network (RDN) as a component with a minor modification and defines it as grouped residual dense block (GRDB). By cascading the GRDBs with attention modules, we could obtain the state-of-the-art performance in real-world image denoising task~\cite{cbam}. We achieved the best performance in terms of the peak signal-to-noise ratio (PSNR) of 39.93 dB and the structural similarity (SSIM) of 0.9736 in the NTIRE2019 Real Image Denoising Challenge - Track 2:sRGB.

%------------------------------------------------------------------------

\section{Related Works}
\subsection{Image Restoration}

Image denoising is one of the most extensively studied topics in image processing. Owing to significant advances in deep learning, CNN-based methods are now dominating in image denoising. However, most previous learning-based image denoising methods have focused on the classic Gaussian denoising task. Toward real-world image denoising, the first approach was to capture a pair of noisy and noise-free images by using different camera settings~\cite{RENOIR,dnd}. It was shown in \cite{dnd} that earlier learning-based methods were comparable or sometimes even inferior to classic methods such as BM3D. We consider this is mainly because of insufficient quality and quantity of training dataset. Consequently, more abundant and elaborate datasets such as Darmstadt noise dataset (DND) and SIDD~\cite{SIDD} were developed, and recent learning-based methods~\cite{SIDD,google_denoier,our_synthetic_dataset,n3net} showed their superiority over classic methods for real-world image denoising.   

In addition to the efforts in generating high quality datasets, a significant amount of research has been made to find better network architectures for image denoising. From the viewpoint of CNNs, network architectures developed for different image restoration tasks such as image denoising, image deblurring, super-resolution, and compress artifact reduction share similarities. It has been repeatedly demonstrated that one architecture developed for a certain image restoration task also performs well in other restoration tasks~\cite{dncnn,rdn2,n3net}. We thus examined many of the architectures developed for different image restoration tasks, especially super-resolution~\cite{srcnn,Kim_sr,laplacian_pyramid_sr,srgan,recursive1,mnet_sr,rdn,rcan,2018srwinner,edsr}. Among them, RDN~\cite{rdn,rdn2} and residual channel attention network (RCAN)~\cite{rcan} are most closely related to our network architecture.

In particular, we attempt to take advantage of novel ideas in RDN and RCAN. RCAN introduced residual in residual (RIR) architecture, and the ablation study showed that the performance gain by RIR was the most significant. Thus, we use this RIR principle in our architecture design. In addition, RDN itself is an image restoration network but we use it with modifications as a component of our network and construct a cascaded structure of RDNs as our image denoising network. Recent studies also showed the effectiveness of attention modules. Among many attention modules, convolutional block attention module (CBAM)~\cite{cbam}, an easily implantable module that sequentially estimates channel attention and spatial attention, showed efficacy in general object detection and image classification, and thus we include CBAM into our network.

\subsection{GAN}
The amount of training images in publicly available real-world image denoising datasets such as SIDD and DND may not be enough to train a deep and wide neural network. One feasible way of augmenting these datasets is to exploit the capability of GAN~\cite{gan}. The first GAN-based real-world noise modeling method~\cite{gan_dataset} uses only real-world noisy images for training the noise generator, where the discriminator is trained to distinguish between real and simulated noise signals. The noise generator is then used to add synthetic but realistic noise to noise-free ground-truth images, and the denoising network is finally trained using the generated pairs of ground-truth and noisy images. The real-world image denoising performance was significantly improved by using the dataset generated by GAN.

We improve the previous GAN-based real-world noise simulation technique~\cite{gan_dataset} by including conditioning signals such as the noise-free image patch, ISO, and shutter speed as additional inputs to the generator. The conditioning on the noise-free image patch can help generating more realistic signal-dependent noise and the other camera parameters can increase controllability and variety of simulated noise signals. We also change the discriminator of the previous architecture~\cite{gan_dataset} by using a recent relativistic GAN~\cite{relativistic_gan}. Unlike conventional GANs, the discriminator of the relativistic GAN learns to determine which is more realistic between real data and fake data. Our method is different from the conventional relativistic GAN in that both the real and fake data are used as an input to make the discriminator more explicitly compare the two data.

\begin{figure}[h]
	\def\arraystretch{0.7}
	\begin{center}
		\def\arraystretch{0.7}
		\begin{tabular}{@{}c@{}}
			\includegraphics[width=1.0\linewidth]{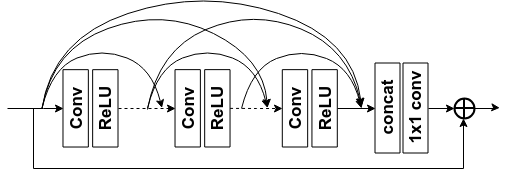} \\ 
			\vspace{0.2cm}
			\small (a) \\ 
			\vspace{0.2cm}
			\includegraphics[width=1.0\linewidth]{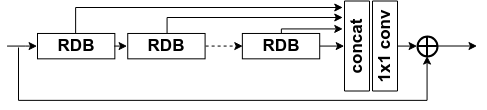} \\
			\vspace{0.2cm}
			\small (b) \\
		\end{tabular}
	\end{center}
	\caption{Components of GRDN: (a) RDB and (b) GRDB.}
	\label{fig:GRDB}
\end{figure}

%------------------------------------------------------------------------
\section{Proposed Methods}
\subsection{Image Denoising Network}

%-------------------------------------------------------------------------

Our image denoising network architecture called GRDN is shown in Fig.~\ref{fig:GRDN}. Our designing principle is to distribute burdens of each layer such that a deeper and wider network can be well trained. To this end, residual connections are applied in four different levels. Down-sampling and up-sampling layers are included to enable a deeper and wider architecture and CBAM~\cite{cbam} is also applied. %The follows describe the details of our network architecture.

Inspired by RDN~\cite{rdn}, we use RDB as shown in Fig.~\ref{fig:GRDB}(a) as a building module. In RDN, the features from cascaded RDBs are concatenated together and followed by the 1${\times}$1 convolutional layer. We define this feature concatenation part of RDN, as shown in Fig.~\ref{fig:GRDB}(b), as GRDB and use it as a building module of our GRDN. Note that the original RDN~\cite{rdn} applies convolutional layers before and after GRDB and uses global residual learning for image denoising. However, we consider that RDN imposes a heavy burden to the very last 1${\times}$1 convolutional layer of GRDB. Therefore, we instead cascade GRDBs such that the features from RDBs can be fused in multiple stages. Motivated by many recent image restoration networks including RDN~\cite{rdn}, we also include the global residual connection such that the network can focus on learning the difference between the noisy and ground-truth images. Last, we exploit CBAM as a building module to further improve the denoising performance. The position of the CBAM block was empirically chosen as in-between the upconvolutional layer and the last convolutional layer.   

\begin{figure}[h]
	\begin{center}
		\includegraphics[width=1.0\linewidth]{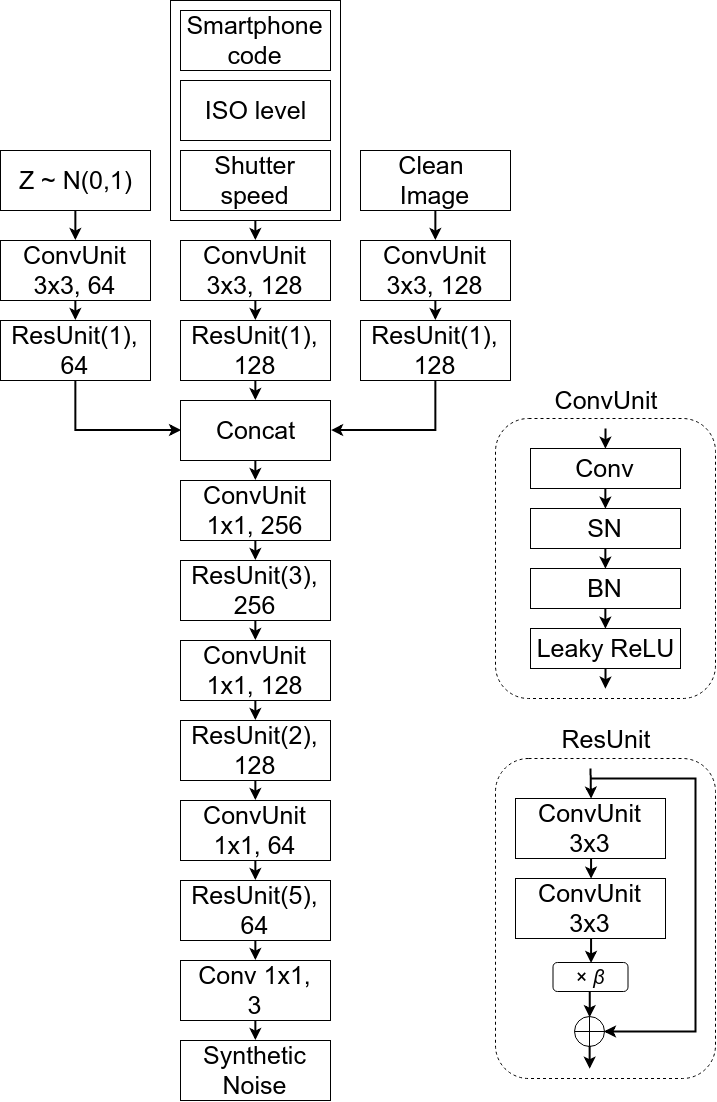}
	\end{center}
	\caption{cERGAN generator.}
	\label{fig:generator}
\end{figure}

Although GRDN is structurally deeper than RDN~\cite{rdn, rdn2}, we used the same number of RDBs. Specifically, 16 RDBs were used in the original RDN for image denoising. We use 4 stack of GRDBs and each GRDB consists of 4 RDBs, resulting 16 RDBs in GRDN.

%------------------------------------------------------------------------

\begin{figure}[h]
	\begin{center}
		\includegraphics[width=1.0\linewidth]{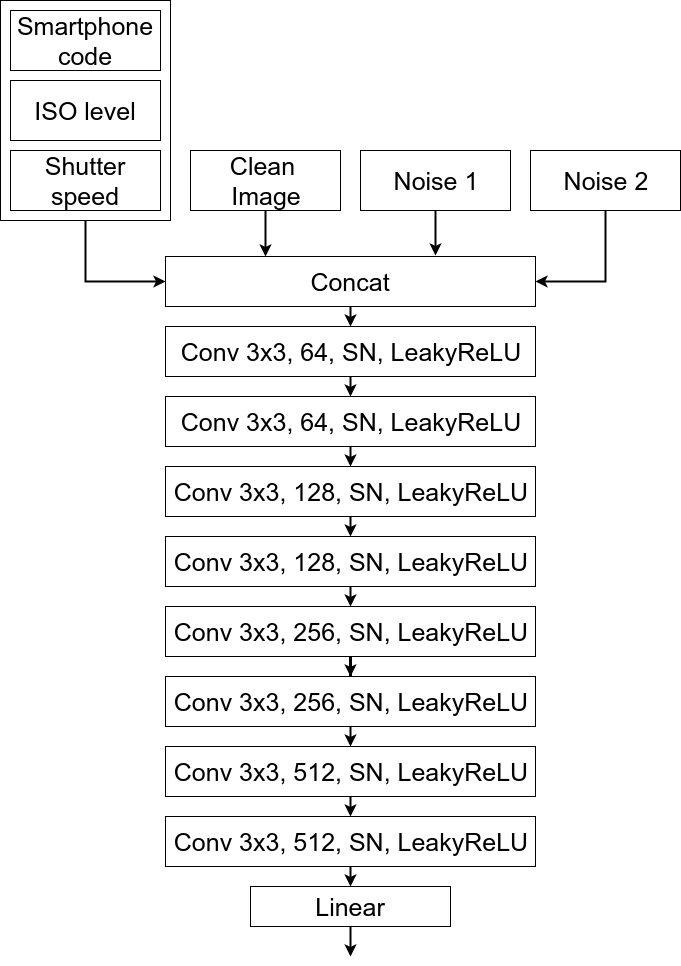}
	\end{center}
	\caption{cERGAN discriminator.}
	\label{fig:discriminator}
\end{figure}

\subsection{GAN-based Real-world Noise Modeling}\label{GAN-based Real-world Noise Modeling}

Motivated by the recent technique~\cite{gan_dataset}, we develop our own generator and discriminator for real-world noise modeling.  
Likewise with the previous technique~\cite{projection_gan}, we use residual blocks (ResBlocks) as a building module of the generator. However, we made several modifications to improve the performance of real-world noise modeling. Fig.~\ref{fig:generator} show the generator architecture. First, we include conditioning signals: the noise-free image patch, ISO, shutter speed, and smartphone model as an additional input to the generator. The conditioning on the noise-free image patch can help generating more realistic signal-dependent noise and the other camera-related parameters can increase controllability and variety of simulated noise signals. To train the generator with these conditioning signals, we used the metadata of SIDD~\cite{SIDD}. Second, spectral normalization (SN)~\cite{sn} is applied before batch normalization in the basic convolutional units like the one used in~\cite{self_attention_gan}. Third, our ResBlock includes the residual scaling~\cite{inception,edsr,esrgan}. SN and residual scaling were empirically found to be useful in training our generator.

Our discriminator architecture as shown in Fig.~\ref{fig:discriminator} is also different from the previous GAN-based noise simulation technique~\cite{gan_dataset}. Enhanced super-resolution GAN (ESRGAN)~\cite{esrgan} showed that relativistic GAN~\cite{relativistic_gan} is effective in generating realistic image textures. Unlike original GAN\cite{gan}, the discriminator of relativistic GAN learns to determine which is more realistic between real data and fake data. Let $C(x)$ denote the non-transformed discriminator output for input image $x$. The standard discriminator can then be expressed as $D(x) = \sigma(C(x))$, $\sigma$ is the sigmoid function. The discriminator of relativistic average GAN (RaGAN) adopted in ESGAN is defined as:
\begin{equation}
\begin{split}
D_{RaGAN}(x_r,x_f)=\sigma(C(x_r)-\E[C(x_f)]), 
\end{split}
\end{equation}
\begin{equation}
\begin{split}
D_{RaGAN}(x_f,x_r)=\sigma(C(x_f)-\E[C(x_r)]), 
\end{split}
\end{equation}
where $x_r$ and $x_f$ denote real data and fake data, respectively, and $\E[\cdot]$ represents the expectation operator, which is applied to all of the data in the mini-batch~\cite{esrgan}.

The discriminator of the proposed network, defined as conditioned explicit relativistic GAN~(cERGAN), is given as 
\begin{equation}
\begin{split}
D_{cERGAN}(x_c,x_r,x_f)=\sigma(C(x_c,x_r,x_f)), 
\end{split}
\end{equation}
\begin{equation}
\begin{split}
D_{cERGAN}(x_c,x_f,x_r)=\sigma(C(x_c,x_f,x_r)), 
\end{split}
\end{equation}
where $x_c$ denote the conditioning signal. Specifically, we make each conditioning data have the same size as the training patch by replicating values, and thus our $x_c$ consists of 4 patches: 3 constant patches from smartphone code (e.g. Google Pixel = 0, iPhone 7 = 1, etc), ISO level, and shutter speed, and one noise-free image patch. In addition to $x_c$, we also use both $x_r$ and $x_f$ as an input of the discriminator. Note that ESGAN uses either $x_r$ or $x_f$ as an input of the discriminator. 

The loss functions of the generator and discriminator, denoted as $L^{cERGAN}_{G}$ and $L^{cERGAN}_{D}$, respectively, are finally defined as follows:
\begin{equation}
\begin{split}
L^{cERGAN}_{G}=\frac{1}{2}\E[(D_{cERGAN}(x_c,x_r,x_f))^2]+\\
\frac{1}{2}\E[(D_{cERGAN}(x_c,x_f,x_r)-1)^2],
\end{split}
\end{equation}
\begin{equation}
\begin{split}
L^{cERGAN}_{D}=\frac{1}{2}\E[(D_{cERGAN}(x_c,x_r,x_f)-1)^2]+\\
\frac{1}{2}\E[(D_{cERGAN}(x_c,x_f,x_r))^2].
\end{split}
\end{equation}
In other words, if the second input is $x_r$ and the third input is $x_f$, the discriminator is trained to predict a value close to 1, i.e., $x_r$ is more realistic than $x_f$. If the two inputs are switched, the discriminator is trained to predict a value close to 0, i.e., $x_f$ is less realistic than $x_r$. The generator is trained to fool the discriminator. By requiring the network to explicitly compare between real data and fake data, we could simulate more realistic real-world noise.

%------------------------------------------------------------------------

\section{Experiments}
We implemented all of our models using
PyTorch library with Intel i7-8700 @3.20GHz, 32GB of
RAM, and NVIDIA Titan XP.

\begin{table*}[!t]
\begin{center}
\begin{tabular}{|c|c|c|c|c|c|c|c|c|c|}
\hline
                & 1st   & 2nd   & 3rd   & 4th    & 5th   & 6th   & 7th   \\ \hline
Model      & RDN   & GRDN  & GRDN  & GRDN   & GRDN  & GRDN  & GRDN  \\ \hline
CBAM            & -     & - &\checkmark&\checkmark&-&\checkmark&\checkmark\\ \hline
Patch size      & 48    & 48    & 48    & 96     &96     &96     & 96    \\ \hline
\# of RDBs      & 16    & 16    & 16    & 16     & 16    & 20    & 16    \\ \hline
\# of filters   & 64    & 64    & 64    & 64     & 64    & 64    & 80    \\ \hline
PSNR (dB)       & 39.37 & 39.41 & 39.46 & 39.62  & 39.63 & 39.65 & \textbf{39.67} \\ \hline
\end{tabular}
\end{center}
\caption{Comparison of image denoising models.}\label{table1}
\end{table*}

\subsection{Datasets}

We used the training and validation images of NTIRE 2019 Real Image Denoising Challenge, which is a subset of SIDD dataset~\cite{SIDD}. Let ChDB denote the dataset we used for our experiment. Specifically, 320 high-resolution images and 1280 cropped image blocks with the size 256$\times$256 were used for training and validation, respectively. The provided images were taken by five smartphone cameras - Apple iPhone 7, Google Pixel, Samsung Galaxy S6 Edge, Motorola Nexus 6, and LG G4. Because the ground-truth images of the test dataset are not publicly available, we report the performance of image denoising models using the validation dataset in this Section. Since we noticed non-marginal degradations around image borders in ground-truth images, we excluded the first and last 8 rows/columns when generating training patches. General data augmentation techniques such as scaling, flipping, and rotation were not applied. 

\subsection{Image Denoising}
\subsubsection{Implementation details}
We augmented the provided training dataset by two ways. First, we used the author-provided source code of \cite{our_synthetic_dataset} for adding synthetic noise to the ground-truth images. We also applied our own GAN-based noise simulator described in Sec.\ref{GAN-based Real-world Noise Modeling} to generate additional synthetic noisy images.

In each training batch, we randomly extracted 16 pairs of ground-truth and noisy image patches. We trained using Adam~\cite{adam} with
${\beta_1}$ = 0.9, ${\beta_2}$ = 0.999. The initial learning rate was set to ${10^{-4}}$ and then decreased to half at every ${2\times10^{5}}$ iteration. We trained the network using $L_1$ loss. We trained our model for approximately 5 days.

We used 4${\times}$4 filters for up/down-convolutional layer and 1${\times}$1 filters for fusing the features concatenated from RDBs. Otherwise, we used 3${\times}$3 filters. Zero-padding was used and dilation was not used for all convolutional layers. Each RDB has 8 pairs of convolutional layers and ReLU activation layers. 

\subsubsection{Comparison to RDN}

First, we compared our GRDN model with RDN~\cite{rdn2}. The experimental result
is shown in Table~\ref{table1}. We re-trained RDN using ChDB. The 1st and 2nd columns in Table~\ref{table1} correspond to RDN and proposed GRDN. It can be seen that the PSNR of our model is 0.04 dB higher than that of RDN. Note that RDN and GRDN have the same number of RDBs, and thus the number of parameters is similar. Specifically, our basic GRDN model has 22M parameters while RDN has 21.9M parameters. %Our model has 0.2\% more parameters than RDN while performance increased by 0.04 dB.

\subsubsection{Experiments on patch size}

Since the original image resolution is very high (more than 12M pixels), the largest possible patch size needs to be used to include sufficient image contents. We thus increased the patch size to 96 $\times$ 96, which was the largest possible size in our experimental environment. 
By comparing the 2nd and 5th columns of Table~\ref{table1}, we can see that the significant performance gain of 0.22dB was obtained by increasing the patch size.

\subsubsection{Experiments on CBAM module}

CBAM~\cite{cbam} is a simple but effective module for CNNs. Because it is a lightweight and general module, it can be easily implanted to any CNN architectures without largely increasing the number of parameters. In particular, CBAM can be placed at bottlenecks of the network. Since we have down-sampling and up-sampling layers, we examined different positions and combinations of CBAMs. We concluded that for our model the best position of CBAM is after the up-sampling layer. We believe this indicates that CBAM enhances important features from the up-sampled data. It also helps to construct a final denoised image for the last convolution layer which comes after. The effectiveness of CBAM was found to be dependent on the complexity of the network.
Comparing the 2nd and 3rd columns of Table~\ref{table1}, CBAM increased the PSNR by 0.05 dB. However, after increasing the patch size, the gain by CBAM became diluted. Comparing the 5th and 6th columns of Table~\ref{table1}, CBAM even decreased the PSNR by 0.01 dB.

\subsubsection{Hyper parameter adjustment}

We compared networks with different numbers of filters and GRDBs. Comparing the 6th and 7th columns of Table~\ref{table1}, a less deeper but more wider network performed 0.02 dB better. Therefore, the model on the 7th column is the best performing model under our hardware constraints.

%-------------------------------------------------------------------------
\subsection{Real-world Noise Modeling}\label{augmentation}

\begin{figure}
 \def\arraystretch{0.7}
  \begin{center}
\def\arraystretch{0.7}
    \begin{tabular}{@{}c@{}}
        \includegraphics[width=1\linewidth]{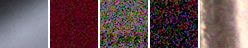} \\
        \vspace{0.1cm}
        \small (a) \\
        \includegraphics[width=1\linewidth]{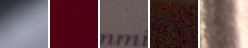} \\ 
        \vspace{0.1cm}
        \small (b) \\
        \includegraphics[width=1\linewidth]{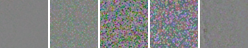} \\ 
        \vspace{0.1cm}
        \small (c) \\
        \includegraphics[width=1\linewidth]{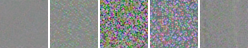} \\ 
        \vspace{0.1cm}
        \small (d) \\
    \end{tabular}
  \end{center}
  \caption{Experimental results on real-world noise modeling: (a) The real-world noisy image patches in ChDB, (b) the ground-truth image patches, (c) difference between noisy and ground-truth image patches, and (d) noise patches generated by cERGAN.}
  \label{fig:gan_results}
\end{figure}

For training the generator and discriminator of cERGAN, we cropped image patches with the size 48$\times$48 from real-world noisy images and their ground-truth images from ChDB. We used the batch size of 32 and Adam optimizer with $\beta_1=0$ and $\beta_2=0.9$. The generator and discriminator were trained for 340k iterations. The initial learning rate was set as 0.0002 for both discriminator and generator, and we linearly decayed the learning rate after 320k iterations such that the learning rate became 0 after the last iteration. Fig.~\ref{fig:gan_results} illustrates some of noise image patches generated by the proposed cERGAN. As can be seen in Figs.~\ref{fig:gan_results}(c) and (d), the proposed cERGAN can generate noise patches close to real-world noise.

\begin{figure}[h]
   \includegraphics[width=1.1\linewidth]{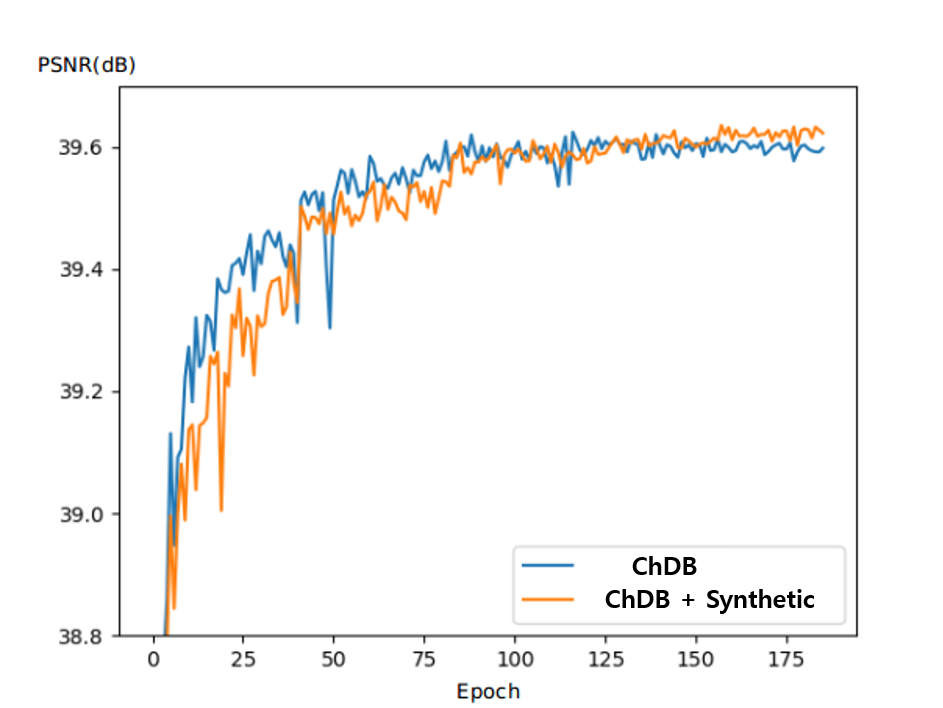}
   \caption{Convergence analysis of image denoising network with different dataset.}
\label{fig:long}
\label{fig:sidd_and_synthetic}
\end{figure}

\begin{table*}[!t]
\begin{center}
\begin{tabular}{@{}ccccc@{}}
\includegraphics[width=0.18\linewidth]{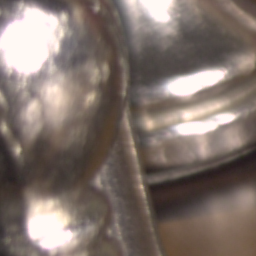}&\includegraphics[width=0.18\linewidth]{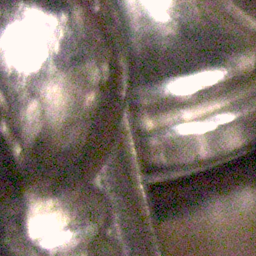}&\includegraphics[width=0.18\linewidth]{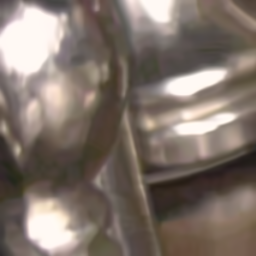}&\includegraphics[width=0.18\linewidth]{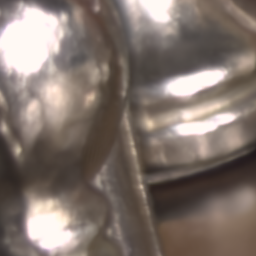}&\includegraphics[width=0.18\linewidth]{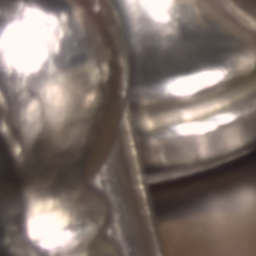}\\
&(26.32 / 0.7576)&(35.49 / 0.9812)&(39.11 / 0.9899)&(39.59 / 0.9902)\\
\includegraphics[width=0.18\linewidth]{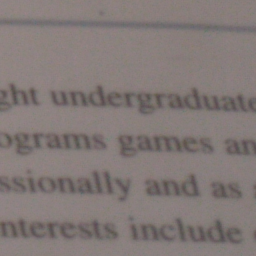}&\includegraphics[width=0.18\linewidth]{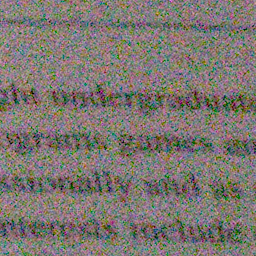}&\includegraphics[width=0.18\linewidth]{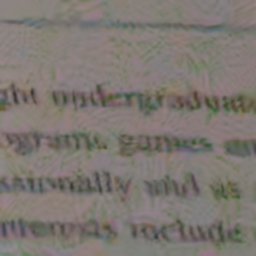}&\includegraphics[width=0.18\linewidth]{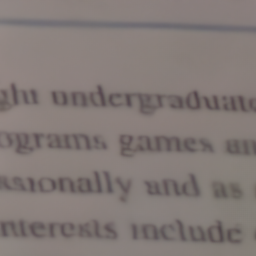}&\includegraphics[width=0.18\linewidth]{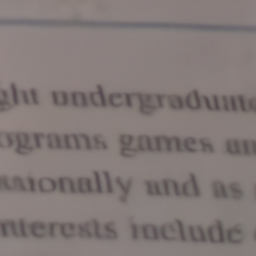}\\
&(19.05 / 0.3623)&(29.86 / 0.9314)&(37.05 / 0.9749)&(37.13 / 0.9748)\\
 \includegraphics[width=0.18\linewidth]{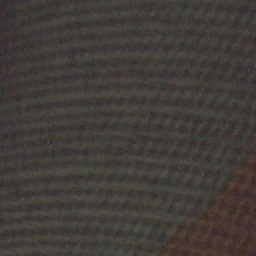}&\includegraphics[width=0.18\linewidth]{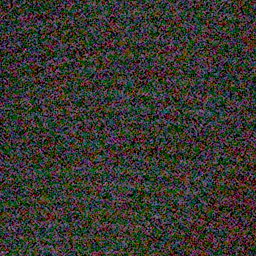}&\includegraphics[width=0.18\linewidth]{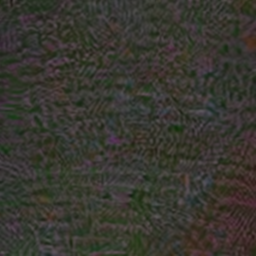}&\includegraphics[width=0.18\linewidth]{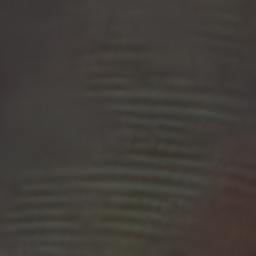}&\includegraphics[width=0.18\linewidth]{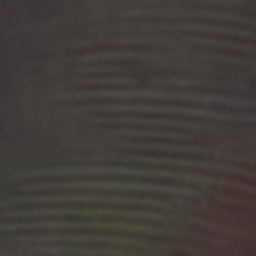}\\
&(17.56 / 0.2444)&(26.27 / 0.8255)&(33.50 / 0.9305)&(33.76 / 0.9347)\\
\includegraphics[width=0.18\linewidth]{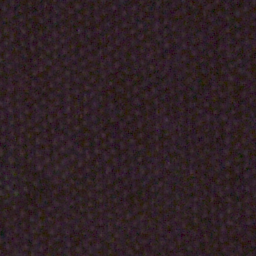}&\includegraphics[width=0.18\linewidth]{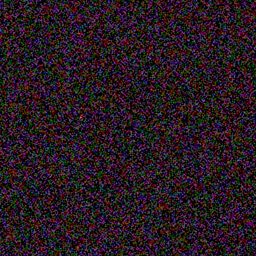}&\includegraphics[width=0.18\linewidth]{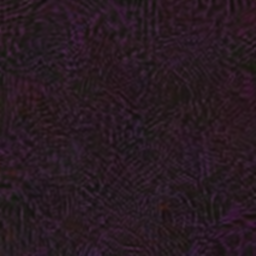}&\includegraphics[width=0.18\linewidth]{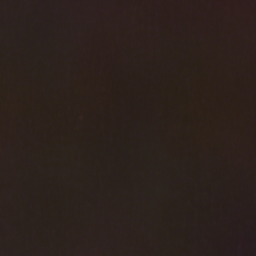}&\includegraphics[width=0.18\linewidth]{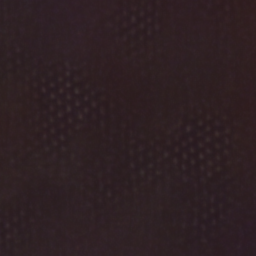}\\
&(18.73 / 0.2757)&(28.12 / 0.8385)&(29.44 / 0.8688)&(31.38 / 0.8846)\\
(a)&(b)&(c)&(d)&(e)\\
\end{tabular}
\captionof{figure}{Comparison of image denoising methods: (a) Ground-truth images, (b) noisy images, (c) denoised images using BM3D with the standard deviation of 50, (d) denoised images using RDN, and (e) denoised images using our proposed network. Below each noisy and denoised image, its quality is provided as (PSNR (dB) / SSIM). The results are best viewed in the electronic version.}
\end{center}
\end{table*}

\begin{table}[]
\centering
{\begin{tabular}{|l|l|l|l|}
\hline
\textbf{Method}                            & \textbf{PSNR}                    & \textbf{SSIM}                   \\ \hline
{\color[HTML]{333333} \textbf{GRDN (Ours)}} & {\color[HTML]{FE0000} \textbf{39.931743}} & {\color[HTML]{FE0000} \textbf{0.973589}} \\ \hline
2nd method                                 & 39.883139                        & 0.973113                        \\ \hline
3rd method                                 & 39.818198                        & 0.972963                        \\ \hline
4th method                                 & 39.675235                        & 0.972554                        \\ \hline
5th method                                 & 39.610533                        & 0.972637                        \\ \hline
\end{tabular}
}
\caption{Performance comparison on the test dataset of NTIRE 2019 Real Image Denoising Challenge - Track2:sRGB.}\label{ntireresulttable}
\end{table}

The effectiveness of simulated noisy images was evaluated by comparing the proposed image denoising network trained with/without the simulated data. Here, the tested network corresponds to the 4th column of Table~\ref{table1}. We first attempted to train our image denoising network using only the synthesized real-world noisy images obtained by cERGAN. The average PSNR was obtained as 38.63 dB in ChDB validation set, which is inferior to the one we obtained using only the provided ChDB dataset (39.62 dB in Table~\ref{table1}).

Second, we used the author-provided source code of \cite{our_synthetic_dataset} for adding statistically modeled real-world noise to ground-truth images of ChDB. Our image denoising network trained using these dataset only resulted in 36.17 dB, which demonstrates that the proposed GAN-based noise modeling at least performs better than the statistic noise modeling method~\cite{our_synthetic_dataset}.

Last, we combined the original ChDB dataset with the synthetic datasets generated by the proposed cERGAN and conventional method~\cite{our_synthetic_dataset}. Here, we could test only one configuration: 90\% from ChDB, 5\% from simulated ChDB using \cite{our_synthetic_dataset}, and 5\% from simulated ChDB using cERGAN.
Fig.~\ref{fig:sidd_and_synthetic} shows that the PSNR obtained using the augmented dataset increases more stably. The resultant PSNR was obtained as 39.64 dB, which is slightly higher than the PSNR obtained using the original dataset (39.62 dB).

%-------------------------------------------------------------------------

\section{NTIRE2019 Image Denoising Challenge}

This work is proposed for participating in the NTIRE2019 Real Image Denoising Challenge - Track 2:sRGB. The challenge aims to develop an image denoising system with the highest PSNR and SSIM. The submitted image denoising network corresponds to 7th column of Table \ref{table1}. One minor change in the submitted model is that we included skip connections for every 2 GRDBs. For training, we used the augmented ChDB using the technique mentioned in Sec.~\ref{augmentation}.
Our model ranked \textbf{1st place} for real image denoising both in terms of PSNR and SSIM. As
shown in Table \ref{ntireresulttable}, our model outperformed the 2nd rank method by 0.05 dB. 


%-------------------------------------------------------------------------

\section{Conclusion}
In this paper, we proposed an improved network architecture for real-world image denoising. By using residual connections extensively and hierarchically, our model achieved the state-of-the-art performance. Furthermore, we developed an improved GAN-based real-world noise modeling method. 

Although we could evaluate the proposed network only to real-world image denoising, we believe that the proposed network is generally applicable. We thus plan to apply the proposed image denoising network to other image restoration tasks. We also could not fully and quantitatively justify the effectiveness of the proposed real-world noise modeling method. A more elaborate design is clearly necessary for better real-world noise modeling. We believe that our real-world noise modeling method can be extended to other real-world degradations such as blur, aliasing, and haze, which will be demonstrated in our future work. 

\section{Acknowledgement}

This work was supported by Institute for Information and communications Technology Promotion(IITP) grant funded by the Korea government(MSIP) 2017-0-00072, Development of Audio/Video Coding and Light Field Media Fundamental Technologies for Ultra Realistic Tera-media.

{\small
\bibliographystyle{ieee_fullname}
\bibliography{egbib}
}

\end{document}